\documentclass[12pt]{article}\usepackage{amssymb,amsmath}
\begin{document}

%%%%%%%%%%%%%%%%%%%%%%%%%%%%%%%%%%%%%%%%%%%%%%%%%%%%%%%%%%%%%%%%%%%%%
%%%%%%%%%%%%%%      TITLE PAGE     %%%%%%%%%%%%%%%%%%%%%%%%%%%%%%%%%%
%%%%%%%%%%%%%%%%%%%%%%%%%%%%%%%%%%%%%%%%%%%%%%%%%%%%%%%%%%%%%%%%%%%%%

\sloppy
\title
%{\hfill{\normalsize\sf FIAN/TD/01-15}    \\
 %           \vspace{1cm}
{\Large  Comparison of two field theoretical forms of gravitational 
wave equations }

\author
 {
       A.I.Nikishov
          \thanks
             {E-mail: nikishov@lpi.ru}
  \\
               {\small \phantom{uuu}}
  \\
           {\it {\small} I.E.Tamm Department of Theoretical Physics,}
  \\
               {\it {\small} P.N.Lebedev Physical Institute, Moscow, Russia}
  \\
  %       {\it {\small} 117924, Leninsky Prospect 53, Moscow, Russia.}
 }
%
%--------------------------------------------------------------------
\maketitle
%--------------------------------------------------------------------
%%%%%%%%%%%%%%%%%%%%%%%%%%%%%%%%%%%%%%%%%%%%%%%%%%%%%%
\begin{abstract}
 In the lowest nonlinear approximation I compare two gravitational wave
 equations,- those of Weinberg and Papapetrou.  The first one is simply
 a form of Einstein equation and the second is claimed to be yet another
 field theoretical form in which the energy-momentum tensor is obtained
 by Belinfante or Rosenfeld method. I show that for interacting
 gravitational field these methods lead to different energy-momentum tensors.
 Both these tensors need to be complemented "by hand" with some interaction
 energy-momentum tensors in order that the conservation laws of the total
 energy-momentum tensor give equation of motion for particles in agreement
 with general relativity. In approximation considered by Thirring, the
 Papapetrou wave equation must coincide with that of Thirring. But they
 differ because Thirring inserted the necessary interaction term. I show
 that Thirring wave equation is equivalent to Weinberg's one. Hence the
 Papapetrou equation is not yet another form of Einstein equation.
 \end{abstract}
\section{Introduction}
 The Einstein equation can be written straightforwardly in a field 
 theoretical form, see equations. (7.6.3) and (7.6.4) in Weinberg's book [1].
 Another form of wave equation was given by Papapetrou [2]. He assumed 
 that $g_{\mu\nu}$ are simply gravitational potentials in flat space and
 both flat space metric $\eta_{\mu\nu}$ and $g_{\mu\nu}$ are present in
 his theory. It is not clear from his paper whether his wave equation is 
 simply another form of Einstein equation or this is another theory
 similar to that of Rosen [3] and others.

 Later, Thirring [4] showed how one can build up general relativity in 
 the lowest nonlinear approximation starting from flat space and using
 field theoretical methods. It is important to note that Thirring 
 chooses one specific way to obtain the nonlinear corrections. Namely, 
 after considering the linear approximation, he switches from $g_{\mu\nu}=
 \eta_{\mu\nu}+h_{\mu\nu}$ to $\sqrt{-g}g^{\mu\nu}\equiv\eta^{\mu\nu}-
 \varphi^{\mu\nu}$.  Only in the linear approximation
 $\varphi^{\mu\nu}$ is equal $\bar h^{\mu\nu}=h^{\mu\nu}-
 \frac12\eta^{\mu\nu}h$. I did not pay attention to this fact and 
 erroneously concluded that Thirring's wave equation do not lead to 
 Schwarzschild solution [5].
 
 Later, Gupta [6], Halpern [7] and others refer to Papapetrou equation
 as a form of Einstein equation. Moreover, Deser [8], following his
 predecessors, claims that he deduces Einstein equation in just a few
 steps by field theoretical means. Yet doubts remain. We note that 
 Halpern [7], following Papapetrou, obtains the the gravitational 
 energy-momentum tensor from general relativity Lagrangian by Belinfante
 method. In the approximation, considered by Thirring, Halpern energy-momentum
 tensor must coincide with Thirring one. It does not where the matter is
 present. Moreover, my calculation shows that for interacting fields the
 Belinfante method leads to an asymmetric energy-momentum tensor. This is a
 good thing for Thirring because to insure the equation of motion for
 particles in agreement with general relativity, he introduces "by hand"
 the asymmetric interaction energy-momentum tensor. Together with 
 gravitational energy-momentum tensor it restores the necessary symmetry.

 On the other hand, Deser [8] uses the Rosenfeld method to obtain the 
 gravitational energy-momentum tensor. He is sure that the result must be
 the same as in Belinfante method. But Rosefeld method gives the
  symmetrical tensor. One might 
 assume that in this method there is no need to insert "by hand" some 
 interaction terms to get the desired equation of motion for particles.
 My calculations show that this is not the case; using the Rosenfeld 
 method one is forced again to insert some interaction terms.

 In view of these doubts it is desirable to obtain an answer to the question:
 what theory is in agreement with general relativity, Thirring theory or
 Papapetrou one? My calculations show that Thirring wave equation is 
 equivalent to Weinberg's one. Hence, Thirring wave equation is in agreement
 with general relativity, not the Papapetrou one.

\section{Conservation laws of total energy-momentum tensor and the equation
of motion for particles}

In this Section we show that the conservation laws, obtained by Thirring in
a linear approximation, continue to hold also in $h^2$ approximation.
We use the notation
$$
 g_{\mu\nu}=\eta_{\mu\nu}+h_{\mu\nu},\quad \eta_{\mu\nu}={\rm diag}(-1,1,1,1),
 \quad u^{\mu}=\frac{dx^{\mu}}{ds}=\dot x^{\mu},\quad \phi_{,\mu}=
 \frac{\partial}{\partial x^{\mu}}\phi.                               \eqno(1)
 $$ 
  
 According to Thirring the total energy-momentum tensor consists of three
  parts:\\
  matter part
  $$
 \stackrel{M}{T^{\mu\nu}}=
\sum_am_au^{\mu}u^{\nu}\frac{ds}{dt}\delta(\vec x-\vec x_a(t)),\quad
u^{\mu}=dx^{\mu}/ds,                  \eqno (2)
$$
 interaction part
 $$
\stackrel{int}{T}{}^{\mu\nu}= 
\stackrel{M}T{}^{\mu\alpha}h_{\alpha}{}^{\nu}, \eqno (3)
 $$
 and gravitational part, consisting of canonical and spin parts
 $$
 t^{\mu\nu}=\stackrel{can}t{}^{\mu\nu}+\stackrel{s}t{}^{\mu\nu}.  \eqno(4)
 $$
 The spin part $\stackrel{s}t{}^{\mu\nu}$ do not contribute to the 
 conservation laws. Calculating the divergence of the canonical tensor and
 using the linearized Einstein equation, we find
 $$
  \stackrel{can}t{}^{\mu\nu}{}_{,\mu}=
  -\frac12h_{\alpha\beta}{}^{,\nu}\stackrel{M}T{}^{\alpha\beta}. \eqno(5)
  $$
 Corrections to the linearized Einstein equation would lead to corrections of
 order $h^3$ in (5) and we neglect them.

 The divergence of matter tensor is
 $$
 \stackrel{M}T{}^{\mu\nu}{}_{,\mu}=
 \sum_am_a\dot u^{\nu}\frac{ds}{dt}\delta(\vec x-\vec x_a(t)),  \eqno(6)
 $$
 see eq.(2.8.6) in [1]. We use here the particle equation of motion in
$h^2$ approximation:
$$
\dot u^{\nu}=-\Gamma_{\alpha\beta}^{\nu}u^{\alpha}u^{\beta}\approx
[-h^{\nu}{}_{\alpha,\beta}+\frac12h_{\alpha\beta}{}^{,\nu}+
h^{\nu\sigma}(h_{\sigma\alpha,\beta}-
\frac12h_{\alpha\beta,\sigma})]u^{\alpha}u^{\beta}. \eqno(7)
$$
Then we get
$$
 \stackrel{M}T{}^{\mu\nu}{}_{,\mu}=
 \sum_am_a\frac{ds}{dt}\delta(\vec x-\vec x_a(t))
 u^{\alpha}u^{\beta}[-h^{\nu}{}_{\alpha,\beta}+\frac12h_{\alpha\beta}{}^{,\nu}+
h^{\nu\sigma}(h_{\sigma\alpha,\beta}-
\frac12h_{\alpha\beta,\sigma})].   \eqno(8)
 $$
For the interaction tensor we have
$$
\stackrel{int}{T}{}^{\mu\nu}{}_{,\mu}=\stackrel{M}T{}^{\mu\sigma}{}_{,\mu}
h_{\sigma}{}^{\nu}+\stackrel{M}T{}^{\mu\sigma}h_{\sigma}{}^{\nu}{}_{,\mu}.
                                                                    \eqno(9)
$$
Using (8), we find with our accuracy
$$
\stackrel{int}{T}{}^{\mu\nu}{}_{,\mu}=
\stackrel{M}T{}^{\mu\sigma}h_{\sigma}{}^{\nu}{}_{,\mu}-
h^{\nu\sigma}(h_{\sigma\alpha,\beta}-
\frac12h_{\alpha\beta,\sigma})]\stackrel{M}{T}{}^{\alpha\beta}.   \eqno(10)
$$
From (8) and (10) we have
$$
 \left(\stackrel{M}T{}^{\mu\nu}+\stackrel{int}T{}^{\mu\nu}
 \right)_{,\mu}=                               
\frac12h_{\alpha\beta}{}^{,\nu}\stackrel{M}T{}^{\alpha\beta}+O(h^3). \eqno(11)
$$
Adding (5), we have the conservation laws for the total energy-momentum tensor
$$
 \left(\stackrel{M}T{}^{\mu\nu}+\stackrel{int}T{}^{\mu\nu}
 +t^{\mu\nu} \right)_{,\mu}=0.                               \eqno(12)
 $$

 It is interesting to note that lowering superscript $\nu$ in (11) with the 
 help of $\eta$, we obtain
 $$
   \left(\stackrel{M}T{}^{\mu}{}_{\nu}+\stackrel{int}T{}^{\mu}{}_{\nu}
 \right)_{,\mu}= \left(\stackrel{M}T{}^{\mu\sigma}(\eta_{\sigma\nu}+
 h_{\sigma\nu}
 \right)_{,\mu}=                               
\frac12h_{\alpha\beta}{}_{,\nu}\stackrel{M}T{}^{\alpha\beta}. \eqno(13)
 $$
 The last equation here is an exact relation of general relativity, see
 eqs. (96.1) and (106.4) in [9]. So in general relativity 
 $\stackrel{M}T{}^{\mu\sigma}(\eta_{\sigma\nu}+
 h_{\sigma\nu})$ is simply a mixed tensor. Only in field theoretical approach
 $\stackrel{int}T{}^{\mu\nu}$ lives in a contravariant form. In general 
 relativity we have only $\stackrel{M}T{}^{\mu\nu}$ and it is a tensor density
 there, see eq.(106.4) in [9].

 \section{Gravitational energy-momentum tensors}

 In this Section we compare gravitational energy-momentum tensors computed
 by Belinfante and Rosenfeld methods. Later we compare also some other
 tensors. To facilitate the comparison, we introduce special notation for
 the building blocks of these tensors:
 $$
 \stackrel{1}{\cal T}{}^{\mu\nu}=\eta^{\mu\nu}\bar h_{\alpha\beta,\gamma}
 \bar h^{\gamma\beta,\alpha};\quad \stackrel{2}{\cal T}{}^{\mu\nu}=
 \eta^{\mu\nu}\bar h_{\alpha\beta,\gamma}\bar h^{\alpha\beta,\gamma};\quad
 \stackrel{3}{\cal T}{}^{\mu\nu}=\eta^{\mu\nu}\bar h_{,\alpha}
 \bar h^{\alpha\beta}{}_{,\beta};\quad\stackrel{4}{\cal T}{}^{\mu\nu}=
 \eta^{\mu\nu}\bar h_{,\alpha}
 \bar h^{,\alpha};\quad
 $$ 
  $$
  \stackrel{16}{\cal T}{}^{\mu\nu}=
 \eta^{\mu\nu}\bar h_{\alpha\lambda}{}^{\lambda}\bar h^{\alpha\sigma}
 {}_{,\sigma};\quad\stackrel{5}{\cal T}{}^{\mu\nu}=\bar h^{\alpha\beta,\mu}
 \bar h_{\alpha\beta}{}^{\nu};\quad \stackrel{6}{\cal T}{}^{\mu\nu}=
 \bar h^{\mu\alpha,\beta}\bar h^{\nu}{}_{\beta,\alpha};\quad
 \stackrel{7}{\cal T}{}^{\mu\nu}=
 \bar h^{\mu\alpha,\beta}\bar h^{\nu}{}_{\alpha,\beta};\quad
 $$
   $$
  \stackrel{8}{\cal T}{}^{\mu\nu}=\frac12(
 \bar h^{\mu\alpha,\nu}+\bar h^{\nu\alpha,\mu})\bar h_{,\alpha};
 \quad\stackrel{9}{\cal T}{}^{\mu\nu}=
 \bar h^{\mu\nu,\alpha}\bar h_{,\alpha};\quad \stackrel{10}{\cal T}{}^{\mu\nu}=
 \frac12(\bar h^{\mu\alpha}{}_{,\alpha}\bar h^{,\nu}+
 \bar h^{\nu\alpha}{}_{,\alpha}\bar h^{,\mu}); \quad
 $$
$$
 \stackrel{11}{\cal T}{}^{\mu\nu}=\bar h^{,\mu}
 \bar h^{,\nu};\quad \stackrel{12}{\cal T}{}^{\mu\nu}=
 \bar h^{\mu\nu,\alpha}
\bar h_{\alpha\beta}{}^{,\beta};\quad\stackrel{13}{\cal T}{}^{\mu\nu}=
 \frac12(\bar h^{\mu\alpha,\beta}\bar h_{\alpha\beta}{}^{,\nu}+
 \bar h^{\nu\alpha,\beta}\bar h_{\alpha\beta}{}^{,\mu}); \quad 
 \stackrel{14}{\cal T}{}^{\mu\nu}=
 \bar h^{\mu\alpha}{}_{,\alpha}\bar h^{\nu\beta}{}_{,\beta};
$$
  $$
\stackrel{15}{\cal T}{}^{\mu\nu}=\frac12(
 \bar h^{\mu\alpha,\nu}+\bar h^{\nu\alpha,\mu})\bar h_{\alpha\beta}{}^{,\beta};
 \quad \bar h^{\mu\nu}=h^{\mu\nu}-
 \frac12\eta^{\mu\nu}h;\quad h=h_{\sigma}{}^{\sigma}=-\bar h.    \eqno(14)
  $$

  Similarly for terms with second derivatives
 $$
\stackrel{a}{\cal T}{}^{\mu\nu}=\eta^{\mu\nu}\bar h_{,\sigma}{}^{\sigma}
\bar h;\quad
\stackrel{b}{\cal T}{}^{\mu\nu}=\eta^{\mu\nu}\bar h_{\alpha\beta}{}^{,\alpha\beta}
\bar h;\quad
\stackrel{c}{\cal T}{}^{\mu\nu}=\eta^{\mu\nu}\bar h^{,\alpha\beta}\bar h_{\alpha\beta};\quad
\stackrel{d}{\cal T}{}^{\mu\nu}=
\eta^{\mu\nu}\bar h^{\alpha\beta,\sigma}{}_{\sigma}\bar h_{\alpha\beta};
$$
$$
\stackrel{e}{\cal T}{}^{\mu\nu}=\eta^{\mu\nu}\bar h^{\alpha\sigma}{}_{,\sigma}{}^{\beta}
\bar h_{\alpha\beta};\quad
\stackrel{f}{\cal T}{}^{\mu\nu}=\bar h^{,\mu\nu}\bar h;\quad 
\stackrel{g}{\cal T}{}^{\mu\nu}=\bar h^{\mu\nu,\sigma}
{}_{\sigma}\bar h;\quad \stackrel{h}{\cal T}{}^{\mu\nu}=
\frac12(\bar h^{\mu\sigma,\nu}{}_{\sigma}+
\bar h^{\nu\sigma,\mu}{}_{\sigma})\bar h;
$$
$$ 
  \quad\stackrel{i}{\cal T}{}^{\mu\nu}=\bar h^{\mu\nu,\alpha\beta}\bar h_{\alpha\beta};\quad
  \stackrel{j}{\cal T}{}^{\mu\nu}=\frac12(\bar h^{\mu\alpha,\nu\beta}+
 \bar h^{\nu\alpha,\mu\beta})\bar h_{\alpha\beta};\quad
 \stackrel{k}{\cal T}{}^{\mu\nu}=
 \bar h^{\alpha\beta,\mu\nu}\bar h_{\alpha\beta};
  \quad\stackrel{l}{\cal T}{}^{\mu\nu}=
 \bar h^{\mu\nu}\bar h_{,\sigma}{}^{\sigma};
  $$
  $$
  \quad\stackrel{m}{\cal T}{}^{\mu\nu}=
 \bar h^{\mu\nu}\bar h_{\alpha\beta}{}^{,\alpha\beta};\quad 
 \stackrel{n}{\cal T}{}^{\mu\nu}=
 \frac12(\bar h^{,\mu\alpha}\bar h_{\alpha}{}^{\nu}+\bar h^{,\nu\alpha}\bar 
 h_{\alpha}{}^{\mu});\quad
 \stackrel{o}{\cal T}{}^{\mu\nu}=
 \frac12(\bar h^{\mu\sigma,\alpha}{}_{\sigma}\bar h_{\alpha}{}^{\nu}+
 \bar h^{\nu\sigma,\alpha}{}_{\sigma}\bar h_{\alpha}{}^{\mu});
  $$
  $$
 \stackrel{p}{\cal T}{}^{\mu\nu}=
 \frac12(\bar h^{\mu\alpha,\sigma}{}_{\sigma}\bar h_{\alpha}{}^{\nu}+
 \bar h^{\nu\alpha,\sigma}{}_{\sigma}\bar h_{\alpha}{}^{\mu});
 \quad\stackrel{q}{\cal T}{}^{\mu\nu}=
 \frac12(\bar h^{\alpha\sigma,\mu}{}_{\sigma}\bar h_{\alpha}{}^{\nu}+
 \bar h^{\alpha\sigma,\nu}{}_{\sigma}\bar h_{\alpha}{}^{\mu}). \eqno(15)
  $$

 Starting from the Lagrangian
 $$
 L=\frac1{32\pi G}[\bar h_{\alpha\beta,\gamma}\bar h^{\gamma\beta,\alpha}-
 \frac12\bar h_{\alpha\beta,\gamma}\bar h^{\alpha\beta,\gamma}+
 \frac14\bar h_{,\sigma}\bar h^{,\sigma}],                           \eqno(16)
 $$
 we find by Belinfante method the following expression for the sum of the
 canonical and spin parts
 $$
 t^{\mu\nu}=\stackrel{can}t{}^{\mu\nu}+\stackrel{s}t{}^{\mu\nu}=
 \frac1{16\pi G}[\frac12\stackrel{1}{\cal T}{}^{\mu\nu}-\frac14
 \stackrel{2}{\cal T}{}^{\mu\nu}+\frac18\stackrel{4}{\cal T}{}^{\mu\nu}+
 \frac12\stackrel{5}{\cal T}{}^{\mu\nu}+\stackrel{6}{\cal T}{}^{\mu\nu}+
 \stackrel{7}{\cal T}{}^{\mu\nu}-\frac14\stackrel{11}{\cal T}{}^{\mu\nu} 
 $$
   $$
   -\stackrel{12}{\cal T}{}^{\mu\nu}-2\stackrel{13}{\cal T}{}^{\mu\nu}
-\stackrel{i}{\cal T}{}^{\mu\nu}+\stackrel{o}{\cal T}{}^{\mu\nu}+
\stackrel{p}{\cal T}{}^{\mu\nu}-\stackrel{q}{\cal T}{}^{\mu\nu}+
   $$
 $$
+\frac12\bar h^{\nu}{}_{\alpha}(\bar h^{\mu\alpha,\sigma}{}_{,\sigma}-
\bar h^{\mu\sigma,\alpha}{}_{,\sigma}-\bar h^{\alpha\sigma,\mu}{}_{,\sigma}-
\frac12\eta^{\mu\alpha}\bar h_{,\sigma}{}^{,\sigma})-
\frac12\bar h^{\mu}{}_{\alpha}(\bar h^{\nu\alpha,\sigma}{}_{,\sigma}-
\bar h^{\nu\sigma,\alpha}{}_{,\sigma}-\bar h^{\alpha\sigma,\nu}{}_{,\sigma}-
\frac12\eta^{\nu\alpha}\bar h_{,\sigma}{}^{,\sigma})]. \eqno(17)
 $$
 Here the antisymmetric part is written down explicitly. Using linearized 
 Einstein equation, we can rewrite it as follows
 $$\frac12(\stackrel{M}{\bar T}{}^{\nu\alpha}\bar h{}_{\alpha}{}^{\mu}-
 \stackrel{M}{\bar T}{}^{\mu\alpha}\bar h{}_{\alpha}{}^{\nu})   \eqno(18)
 $$
 In this expression we can drop bars over $h$ and $\stackrel{M}T$. Then 
 we see that (17), where the antisymmetric part is approximately equal (18),
 together with interaction term (3) yields the symmetric tensor. 

 At this stage it is convenient to compare  (17) with the corresponding 
 Halpern result. Barring a few misprints, the symmetric part in (17)
  agrees with  eq. (3.5a) in [7].
 Instead of our antisymmetric part, Halpern gives (without comments)
  the following asymmetric part
 $$
 \frac1{32\pi G}\bar h^{\mu}{}_{\alpha}(\bar h^{\nu\alpha,\sigma}{}_{,\sigma}-
\bar h^{\nu\sigma,\alpha}{}_{,\sigma}-\bar h^{\alpha\sigma,\nu}{}_{,\sigma}-
\frac12\eta^{\nu\alpha}\bar h_{,\sigma}{}^{,\sigma}), 
 $$
see eq. (3.5b) in [7].
 Turning back to (17), we note that using linearized Einstein equation, we
 find
 $$
 \frac1{16\pi G}\stackrel{p}{\cal T}{}^{\mu\nu}=-\frac12
 (\stackrel{M}{\bar T}{}^{\nu\alpha}\bar h{}_{\alpha}{}^{\mu}+
 \stackrel{M}{\bar T}{}^{\mu\alpha}\bar h{}_{\alpha}{}^{\nu})+
 \frac1{16\pi G}(\frac12\stackrel{l}{\cal T}{}^{\mu\nu}+
 \stackrel{o}{\cal T}{}^{\mu\nu}+\stackrel{q}{\cal T}{}^{\mu\nu}).
                                                                    \eqno(19)
 $$
 Similarly we get
 $$
 \frac1{16\pi G}(\frac12\stackrel{l}{\cal T}{}^{\mu\nu}+
 \stackrel{m}{\cal T}{}^{\mu\nu})=-\frac12\bar h^{\mu\nu}\stackrel{M}T,
 \quad\stackrel{M}T=\stackrel{M}T{}_{\sigma}{}^{\sigma}.        \eqno(20)
 $$
 So we have        
 $$
 \frac1{16\pi G}\stackrel{p}{\cal T}{}^{\mu\nu}=-\frac12
 (\stackrel{M}{\bar T}{}^{\nu\alpha}\bar h{}_{\alpha}{}^{\mu}+
 \stackrel{M}{\bar T}{}^{\mu\alpha}\bar h{}_{\alpha}{}^{\nu})+
 \frac1{16\pi G}(-\stackrel{m}{\cal T}{}^{\mu\nu}+
 \stackrel{o}{\cal T}{}^{\mu\nu}+ 
 \stackrel{q}{\cal T}{}^{\mu\nu})
 -\frac12\bar h^{\mu\nu}\stackrel{M}T.
                                                                    \eqno(21)
  $$
  Now the sum of the r.h.s. of (21) and (18) yields
  $$
  -\stackrel{M}{\bar T}{}^{\mu\alpha}\bar h{}_{\alpha}{}^{\nu}+
 \frac1{16\pi G}(\stackrel{q}{\cal T}{}^{\mu\nu}+
 \stackrel{o}{\cal T}{}^{\mu\nu}-\stackrel{m}{\cal T}{}^{\mu\nu})
 -\frac12\bar h^{\mu\nu}\stackrel{M}T.                        \eqno(22)
 $$
 The first term here is
 $$
 -\bar h^{\nu}{}_{\alpha}(\stackrel{M}T{}^{\mu\alpha}-\frac12\eta^{\mu\alpha}
 \stackrel{M}T{})=
   -\bar h^{\nu}{}_{\alpha}\stackrel{M}T{}^{\mu\alpha}+\frac12\bar h^{\mu\nu}
   \stackrel{M}T{}.                                                 \eqno(23)
 $$
 So the last term in (22) is cancelled by the last term in (23).
 The first term in (23) has the form   
 $$
 -(h^{\nu}{}_{\alpha}-\frac12\eta^{\nu}{}_{\alpha}h)\stackrel{M}T{}^{\mu\alpha}=
 -\stackrel{int}T{}^{\mu\nu}+\frac12h\stackrel{M}T{}^{\mu\nu}.      \eqno(24)
 $$
 Collecting all these results, we finally obtain
   $$
 \stackrel{int}T{}^{\mu\nu}+t^{\mu\nu}= \frac1{16\pi G}[\frac12\stackrel{1}{\cal T}{}^{\mu\nu}-\frac14
 \stackrel{2}{\cal T}{}^{\mu\nu}+\frac18\stackrel{4}{\cal T}{}^{\mu\nu}+
 \frac12\stackrel{5}{\cal T}{}^{\mu\nu}+\stackrel{6}{\cal T}{}^{\mu\nu}+
 \stackrel{7}{\cal T}{}^{\mu\nu}-\frac14\stackrel{11}{\cal T}{}^{\mu\nu} 
 $$
   $$
   -\stackrel{12}{\cal T}{}^{\mu\nu}-2\stackrel{13}{\cal T}{}^{\mu\nu}
-\stackrel{i}{\cal T}{}^{\mu\nu}-\stackrel{m}{\cal T}{}^{\mu\nu}+
2\stackrel{o}{\cal T}{}^{\mu\nu}]+\frac12h\stackrel{M}T{}^{\mu\nu}. \eqno(25)
   $$
   Together with $\stackrel{M}T{}^{\mu\nu}$ it gives the total energy-momentum
   tensor. We shall see later that it leads to the agreement of Thirring wave
    equation with general relativity,

    Now we give the energy-momentum tensor, obtained from the same 
    Lagrangian (16) by Rosenfeld method:
    $$
    t_{Ros}^{\mu\nu}= \frac1{16\pi G}[\frac12\stackrel{1}{\cal T}{}^{\mu\nu}-\frac14
 \stackrel{2}{\cal T}{}^{\mu\nu}+\frac18\stackrel{4}{\cal T}{}^{\mu\nu}+
 \frac12\stackrel{5}{\cal T}{}^{\mu\nu}+\stackrel{6}{\cal T}{}^{\mu\nu}+
 \stackrel{7}{\cal T}{}^{\mu\nu}-\frac14\stackrel{11}{\cal T}{}^{\mu\nu} 
 $$
    $$
   -\stackrel{12}{\cal T}{}^{\mu\nu}-2\stackrel{13}{\cal T}{}^{\mu\nu}
-\stackrel{i}{\cal T}{}^{\mu\nu}+\frac12\stackrel{l}{\cal T}{}^{\mu\nu}+
2\stackrel{o}{\cal T}{}^{\mu\nu}].                                   \eqno(26)
   $$
 Using (20) we find for the difference of (25) and (26):
 $$
 \stackrel{int}T{}^{\mu\nu}+t^{\mu\nu}-t_{Ros}^{\mu\nu}=
 \frac12h\stackrel{M}T{}^{\mu\nu}+\frac12\bar h^{\mu\nu}\stackrel{M}T.\eqno(27)
 $$
 Its divergence is not zero. Thus, $t_{Ros}^{\mu\nu}$ also needs
  some interaction terms.

  Deser  uses the method in which both $\sqrt{-g}g^{\mu\nu}$ and 
  $\Gamma^{\sigma}_{\mu\nu}$  are a priory independent. He calculates 
  energy- momentum tensor by Rosefeld method. It seems that his method must
  be equivalent the usual method in which $\Gamma^{\sigma}_{\mu\nu}$ 
  are functions of metric. If so, his wave equation is not the Einstein one
  already in the considered here approximation.

 \section{Equivalence of Thirring and Weinberg wave equations}

 In terms of $\bar h_{\mu\nu}=h_{\mu\nu}-
 \frac12\eta_{\mu\nu}h$ where $g_{\mu\nu}\equiv
 \eta_{\mu\nu}+h_{\mu\nu}$, the Weinberg form of the Einstein equations is
 $$
 \bar h_{\mu\nu,\sigma}{}^{\sigma}-(\bar h_{\mu\sigma,\nu}{}^{\sigma}+
 \bar h_{\nu\sigma,\mu}{}^{\sigma})+\eta_
 {\mu\nu}\bar h_{\alpha\beta}{}^{\alpha\beta}=-16\pi G(T_{\mu\nu}+
 t^{(2)}_{\mu\nu}).                \eqno(28)
 $$
 Here $t^{(2)}_{\mu\nu}$ is given below, see (34). Thirring deals with
 $\varphi^{\mu\nu}$, defined by
 $$
 \sqrt{-g}g^{\mu\nu}\equiv\eta^{\mu\nu}-
 \varphi^{\mu\nu}.                        \eqno(29)
 $$
 It seems more natural to remain with $h_{\mu\nu}$.
  Then his approach would lead to disagreement with general relativity.    
 Thirring succeeded in obtaining the correct expression for the Schwarzschild 
 solution in the considered approximation. To be sure that there is a 
 complete agreement with general relativity, we have to show that his wave 
 equation is equivalent to Weinberg's one. To that end we rewrite (28) in
 terms of $\varphi_{\mu\nu}=
 \eta_{\mu\alpha}\eta_{\nu\beta}\varphi^{\alpha\beta}$. With our accuracy 
 from definition (29) we have
$$
\varphi^{\mu\nu}=\bar h^{\mu\nu}-\eta^{\mu\nu}(\frac18\bar h^2-
\frac14\bar h_{\alpha\beta}\bar h^{\alpha\beta})+\frac12\bar h^{\mu\nu}\bar h-
\bar h^{\mu}{}_{\tau}\bar h^{\tau\nu}.                              \eqno(30)
$$
In linear approximation $\varphi^{\mu\nu}=\bar h^{\mu\nu}$  and in quadratic 
terms we can substitute $\varphi^{\mu\nu}\leftrightarrow\bar h^{\mu\nu}$.
So we get
$$
\bar h^{\mu\nu}=\varphi^{\mu\nu}+\eta^{\mu\nu}(\frac18\varphi^2-
\frac14\varphi_{\alpha\beta}\varphi^{\alpha\beta})-
\frac12\varphi^{\mu\nu}\varphi+
\varphi^{\mu}{}_{\tau}\varphi^{\tau\nu}.                            \eqno(31)
$$
Lowering here the superscripts $\mu$ and $\nu$ with the help of $\eta$ and 
inserting into (28), we find
 $$
 \varphi_{\mu\nu,\sigma}{}^{\sigma}-(\varphi_{\mu\sigma,\nu}{}^{\sigma}+
 \varphi_{\nu\sigma,\mu}{}^{\sigma})+\eta_
 {\mu\nu}\varphi_{\alpha\beta}{}^{\alpha\beta}=-16\pi G(T_{\mu\nu}+
 t^{(2)}_{\mu\nu})+[-\stackrel{1}{\cal T}{}_{\mu\nu}+
 \stackrel{2}{\cal T}{}_{\mu\nu}+\stackrel{3}{\cal T}{}_{\mu\nu}-                
 $$
  $$
 \frac12\stackrel{4}{\cal T}{}_{\mu\nu}-\stackrel{5}{\cal T}{}_{\mu\nu}-
 2\stackrel{7}{\cal T}{}_{\mu\nu}-\stackrel{8}{\cal T}{}_{\mu\nu} +
 \stackrel{9}{\cal T}{}_{\mu\nu}-\stackrel{10}{\cal T}{}_{\mu\nu}
 +\frac12\stackrel{11}{\cal T}{}_{\mu\nu}+2\stackrel{13}{\cal T}{}_{\mu\nu}
 +2\stackrel{15}{\cal T}{}_{\mu\nu}-\stackrel{16}{\cal T}{}_{\mu\nu}]+
  $$
$$
\{-\frac12\stackrel{a}{\cal T}{}_{\mu\nu}+
\frac12\stackrel{b}{\cal T}{}_{\mu\nu}+
\frac12\stackrel{c}{\cal T}{}_{\mu\nu}+\stackrel{d}{\cal T}{}_{\mu\nu}-
2\stackrel{e}{\cal T}{}_{\mu\nu}+\frac12\stackrel{f}{\cal T}{}_{\mu\nu}+
\frac12\stackrel{g}{\cal T}{}_{\mu\nu}-\stackrel{h}{\cal T}{}_{\mu\nu}+
2\stackrel{j}{\cal T}{}_{\mu\nu}-\stackrel{k}{\cal T}{}_{\mu\nu} +
\frac12\stackrel{l}{\cal T}{}_{\mu\nu}-
$$
$$
\stackrel{n}{\cal T}{}_{\mu\nu} -
2\stackrel{p}{\cal T}{}_{\mu\nu}+2\stackrel{q}{\cal T}{}_{\mu\nu}\}. \eqno(32)
$$
   
Now we write down the Weinberg tensor $t^{(2)}_{\mu\nu}$ in our notation
$$
-16\pi Gt^{(2)}_{\mu\nu}=[\frac12\stackrel{1}{\cal T}{}_{\mu\nu}-
 \frac34\stackrel{2}{\cal T}{}_{\mu\nu}-\stackrel{3}{\cal T}{}_{\mu\nu}                
 $$
  $$
 +\frac38\stackrel{4}{\cal T}{}_{\mu\nu}+
 \frac12\stackrel{5}{\cal T}{}_{\mu\nu}-\stackrel{6}{\cal T}{}_{\mu\nu}+
 \stackrel{7}{\cal T}{}_{\mu\nu}+\stackrel{8}{\cal T}{}_{\mu\nu} -
 \stackrel{9}{\cal T}{}_{\mu\nu}+\stackrel{10}{\cal T}{}_{\mu\nu}
 -\frac14\stackrel{11}{\cal T}{}_{\mu\nu}+\stackrel{12}{\cal T}{}_{\mu\nu}
 -2\stackrel{15}{\cal T}{}_{\mu\nu}
 +\stackrel{16}{\cal T}{}_{\mu\nu}]+
  $$
$$
\{+\frac12\stackrel{a}{\cal T}{}_{\mu\nu}-
\frac12\stackrel{b}{\cal T}{}_{\mu\nu}-
\frac12\stackrel{c}{\cal T}{}_{\mu\nu}-\stackrel{d}{\cal T}{}_{\mu\nu}+
2\stackrel{e}{\cal T}{}_{\mu\nu}-\frac12\stackrel{f}{\cal T}{}_{\mu\nu}-
\frac12\stackrel{g}{\cal T}{}_{\mu\nu}+\stackrel{h}{\cal T}{}_{\mu\nu}+
\stackrel{i}{\cal T}{}_{\mu\nu}-
2\stackrel{j}{\cal T}{}_{\mu\nu}+\stackrel{k}{\cal T}{}_{\mu\nu}- 
$$
$$
\frac12\stackrel{l}{\cal T}{}_{\mu\nu}-
\stackrel{m}{\cal T}{}_{\mu\nu} +
\stackrel{n}{\cal T}{}_{\mu\nu}\}.                              \eqno(33)
$$
From (32)  and (33) we have
 $$
 \varphi_{\mu\nu,\sigma}{}^{\sigma}-(\varphi_{\mu\sigma,\nu}{}^{\sigma}+
 \varphi_{\nu\sigma,\mu}{}^{\sigma})+\eta_
 {\mu\nu}\varphi_{\alpha\beta}{}^{\alpha\beta}=-16\pi GT_{\mu\nu}+
 \{\stackrel{i}{\cal T}{}_{\mu\nu}-\stackrel{m}{\cal T}{}_{\mu\nu}- 
2\stackrel{p}{\cal T}{}_{\mu\nu}+2\stackrel{q}{\cal T}{}_{\mu\nu}\}
$$
  $$
  [-\frac12\stackrel{1}{\cal T}{}_{\mu\nu}+
 \frac14\stackrel{2}{\cal T}{}_{\mu\nu}                
 -\frac18\stackrel{4}{\cal T}{}_{\mu\nu}-
 \frac12\stackrel{5}{\cal T}{}_{\mu\nu}-\stackrel{6}{\cal T}{}_{\mu\nu}-
 \stackrel{7}{\cal T}{}_{\mu\nu}+\frac14\stackrel{11}{\cal T}{}_{\mu\nu}+
 \stackrel{12}{\cal T}{}_{\mu\nu}
 +2\stackrel{13}{\cal T}{}_{\mu\nu}].                              \eqno(34)
  $$

  Next we have to express $T_{\mu\nu}$ in terms of $\stackrel{M}T{}^{\mu\nu}$.
  From definition of $T_{\mu\nu}$ we have
  $$
 T_{\mu\nu}=g_{\mu\alpha}g_{\nu\beta}(-g)^{-1/2}\stackrel{M}T{}^{\alpha\beta}
 \approx(1-\frac12h)\stackrel{M}T{}_{\mu\nu}+
 (\stackrel{M}T{}_{\mu}{}^{\beta}h_{\beta\nu}+
 \stackrel{M}T{}_{\nu}{}^{\beta}h_{\beta\mu}),\quad g={\rm det g_{\mu\nu}}
 \approx -(1+h).                     \eqno(35)
 $$
 In terms of $\bar h_{\mu\nu}$ we have
 $$
 T_{\mu\nu}=\stackrel{M}T_{\mu\nu}+
 (\stackrel{M}T_{\mu}{}^{\beta}\bar h_{\beta\nu}+
 \stackrel{M}T_{\nu}{}^{\beta}\bar h_{\beta\mu})-
 \frac12\stackrel{M}T_{\mu\nu}\bar h.                             \eqno(36)
 $$
 Using linearized Einstein equation
 $$
  \bar h_{\mu\nu,\sigma}{}^{\sigma}-(\bar h_{\mu\sigma,\nu}{}^{\sigma}+
 \bar h_{\nu\sigma,\mu}{}^{\sigma})+\eta_
 {\mu\nu}\bar h_{\alpha\beta}{}^{\alpha\beta}=-16\pi GT_{\mu\nu},
 $$
  we find
 $$
-\frac12\stackrel{M}T_{\mu\nu}\bar h=
\frac1{16\pi G}[\frac12\stackrel{b}{\cal T}{}_{\mu\nu}+
\frac12\stackrel{g}{\cal T}_{\mu\nu}-\stackrel{h}{\cal T}_{\mu\nu}]=
\frac12\stackrel{M}T_{\mu\nu}h,                                  \eqno(37)
 $$
 and
 $$
 \stackrel{M}T_{\mu}{}^{\beta}\bar h_{\beta\nu}+
 \stackrel{M}T_{\nu}{}^{\beta}\bar h_{\beta\mu}=-
 \frac1{16\pi G}[2\stackrel{m}{\cal T}_{\mu\nu}-
2\stackrel{o}{\cal T}_{\mu\nu}+2\stackrel{p}{\cal T}_{\mu\nu}-
2\stackrel{q}{\cal T}_{\mu\nu}].                               \eqno(38)
$$
So eq.(34) takes the form 
$$
\varphi_{\mu\nu,\sigma}{}^{\sigma}-(\varphi_{\mu\sigma,\nu}{}^{\sigma}+
 \varphi_{\nu\sigma,\mu}{}^{\sigma})+\eta_
 {\mu\nu}\varphi_{\alpha\beta}{}^{\alpha\beta}=
 $$
 $$
 -16\pi GT_{\mu\nu}+
 \{\stackrel{i}{\cal T}{}_{\mu\nu}+\stackrel{m}{\cal T}{}_{\mu\nu}- 
2\stackrel{o}{\cal T}{}_{\mu\nu}-\frac12\stackrel{b}{\cal T}{}_{\mu\nu}
-\frac12\stackrel{g}{\cal T}{}_{\mu\nu}+\stackrel{h}{\cal T}{}_{\mu\nu}\}
$$
  $$
  [-\frac12\stackrel{1}{\cal T}{}_{\mu\nu}+
 \frac14\stackrel{2}{\cal T}{}_{\mu\nu}                
 -\frac18\stackrel{4}{\cal T}{}_{\mu\nu}-
 \frac12\stackrel{5}{\cal T}{}_{\mu\nu}-\stackrel{6}{\cal T}{}_{\mu\nu}-
 \stackrel{7}{\cal T}{}_{\mu\nu}+\frac14\stackrel{11}{\cal T}{}_{\mu\nu}+
 \stackrel{12}{\cal T}{}_{\mu\nu}
 +2\stackrel{13}{\cal T}{}_{\mu\nu}].                              \eqno(39)
  $$
 This is Weinberg equation in terms of $\varphi_{\mu\nu}$. 
 As mentioned earlier,
 we may use $\varphi_{\mu\nu}=\bar h_{\mu\nu}$ in quadratic terms. Raising 
 in (39) $\mu$ and $\nu$ with the help of $\eta$, we obtain the wave equation 
 in Thirring form:
 $$
 \varphi^{\mu\nu,\sigma}{}_{\sigma}-(\varphi^{\mu\sigma,\nu}{}_{\sigma}+
 \varphi^{\nu\sigma,\mu}{}_{\sigma})+
 \eta^{\mu\nu}\varphi_{\alpha\beta}{}^{\alpha\beta}=-16\pi G
 (\stackrel{M}T{}^{\mu\nu}+\stackrel{int}T{}^{\mu\nu}+t^{\mu\nu}),    \eqno(40)
 $$
 see eqs, (76) and (77) in [4]. Here use has been made of eqs. (25) and (37).
 Thus Thirring wave equation agrees with general relativity.
   
 \section{Conclusion}
 It is shown that Thirring's field theoretical method to reproduce 
 general relativity in the lowest nonlinear approximation is successful.
 Yet his approach is specific. More natural way leads to failure. There
 are no rigorous proofs that field theoretical derivations of gravitational
 wave equation lead to exact Einstein equation. These facts should stimulate
 the search for alternative theories of gravity, hopefully without black
 holes. The latter are undesirable on energy grounds [6,10]. One particular
 version of gravitational theory without black holes is elaborated by
 Logunov and his colleagues [11]. Another possibility is considered in [12].

For me it seems natural to assume that in three graviton vertex each
graviton interacts with energy-momentum tensor formed by other two
gravitons.One promising way to build up a gravitation theory is to deal
directly with $S-$matrix formalism and use vertices and free propagatores
without any recourse to a wave equation. 
\section{Acknowledgements}
I am grateful to V.I.Ritus and R.Metsaev for valuable discussions and help.   
 This work was supported in part by the Russian Foundation for
Basic Research (projects no. 00-15-96566 and 02-02-16944).

 \section*{References}

1 . S.Weinberg,{\sl Gravitation and Cosmology}, New York (1972).\\
2. A.Papapetrou, Proc. Roy. Irish. Acad. {\bf52A}, 11 (1948).\\
3. N.Rosen, Phys.Rev. {\bf57,}147 (1940).\\
4. W.E.Thirring, Ann. Phys. (N.Y.) {\bf16}, 96 (1961).\\  
5. A.Nikishov, gr-qc/9912034; Part. and Nucl. {\bf32}, 5 (2001).\\
6. S.Gupta, Proc. Phys. Soc. {\bf A 65}, 608 (1952).\\
7. L.Halpern, Bulletin de l'Academie royal de Belgique, Classe de Sciences,
   {\bf49}, 256 (1963).\\
8 S.Deser, Gen. Rel. and Grav., {\bf1}, 9 (1970).\\
9.L.D.Landau and E.M.Lifshitz, {\sl The classical theory of
 fields}, Moscow, (1973) (in Russian).\\
 10. A.Nikishov, gr-qc/0310072.\\
 11. A.A.Logunov, Part. and Nucl. {\bf29} (1) Jan.-Feb 1998, p 1;
{\sl The Theory of Gravitational Field}, Moscow, Nauka (2000), (in Russian). \\
 12.Yu.Baryshev, gr-qc/9912003.
\end{document}